\documentclass[]{raa}            

\input{psfig.sty}
\input{epsf.sty}
\usepackage{graphicx,times}
\usepackage{natbib}

\def\beq{\begin{equation}}
\def\enq{\end{equation}}

\begin{document}

   \title{Possible Distance Indicators in Gamma-ray Pulsars}

   \volnopage{Vol.0 (200x) No.0, 000--000}      
   \setcounter{page}{1}          

   \author{W. Wang
   }

   \institute{National Astronomical Observatories, Chinese Academy of Sciences,
             Beijing 100012, China; {\it wangwei@bao.ac.cn}}
   \date{Received~~2011 month day; accepted~~2011~~month day}

\abstract{ Distance measurements of gamma-ray pulsars are
challenging questions in present pulsar studies. The Large Area
Telescope (LAT) aboard the Fermi gamma-ray observatory discovered
more than 70 gamma-ray pulsars including 24 new gamma-selected
pulsars which nearly have no distance information. We study the
relation between gamma-ray emission efficiency
($\eta=L_\gamma/\dot E$) and pulsar parameters for young
radio-selected gamma-ray pulsars with known distance information
in the first gamma-ray pulsar catalog reported by Fermi/LAT. We
have introduced three generation order parameters to describe
gamma-ray emission properties of pulsars, and find the strong
correlation of $\eta-\zeta_3$ a generation order parameter which
reflects $\gamma$-ray photon generations in pair cascade processes
induced by magnetic field absorption in pulsar magnetosphere. A
good correlation of $\eta-B_{\rm LC}$ the magnetic field at the
light cylinder radius is also found. These correlations would be
the distance indicators in gamma-ray pulsars to evaluate distances
for gamma-selected pulsars. Distances of 25 gamma-selected pulsars
are estimated, which could be tested by other distance measurement
methods. Physical origin of the correlations may be also
interesting for pulsar studies. }

\authorrunning{W. Wang }            
   \titlerunning{Distance Indicators in Gamma-ray Pulsars}

\maketitle

\section{Introduction}

Before 2008, only 7 gamma-ray pulsars are known in nature
(Thompson 2001). The launch of the {Fermi Gamma-ray Space
Observatory} in June 2008 completely changed the status in studies
of gamma-ray pulsars. The first published catalog of gamma-ray
pulsars (Abdo et al. 2010) contains 46 gamma-ray pulsars including
8 millisecond pulsars, 21 young radio pulsars and 17
gamma-selected pulsars. After more than one and half years of
all-sky survey observations by Fermi/LAT, more than 70 gamma-ray
pulsars were discovered, including 25 gamma-selected pulsars (see
reviews by Ray \& Saz Parkinson 2010). High sensitivity of the
Fermi/LAT makes a new era for pulsar discoveries, specially for
the population of radio-quiet gamma-ray pulsars.

The distance measurement of pulsars is always a difficult problem
in pulsar studies. Trigonometric parallax measurements of radio
pulsars are the reliable method, but are only available for the
nearby pulsars ($<0.4$ kpc) specially for a few radio millisecond
pulsars (e.g. Lommen et al. 2006). The most common way to obtain
radio pulsar distance is based on the computation from dispersion
measurement (DM) coupled to an electron density distribution model
like NE 2001 model (Cordes \& Lazio 2002), which have been applied
to most radio pulsars (e.g., Johnston et al. 1996; Keith et al.
2008). The pulsar distance can be also estimated from kinematic
model: the distance of possible associated objects (supernova
remnants, pulsar wind nebulae, star clusters, or HII regions)
could be measured from Doppler shift of absorption or emission
lines in HI spectrum together with the rotation curve model of the
Galaxy (e.g., Robert et al. 1993; Camilo et al. 2006). The
distance of some pulsars with X-ray emissions can be estimated
from X-ray observations of the absorbing column (e.g. Romoni et
al. 2005) or from correlations in X-ray luminosities versus
spin-down power or photon index (Becker \& Truemper 1997; Possenti
et al. 2002; Gotthelf 2003; Wang 2009; and references therein).
These methods may be available for radio or even X-ray pulsars,
but for gamma-selected pulsars if no possible associated objects,
we would have no any information on their distance.

It is well known that X-ray luminosity has the correlation with
pulsar's spin-down power: $L_x\propto \dot E$ in soft X-ray bands
(0.1 -- 2.4 keV, Becker \& Truemper 1997), and $L_x\propto \dot
E^{3/2}$ in hard X-ray bands ($>2$ keV, see Saito 1998; Cheng et
al. 2004; Wang 2009). Based on the EGRET pulsars, Thompson et al.
(1999) found a possible correlation of $L_\gamma\propto \dot
E^{1/2}$. For the larger sample of gamma-ray pulsars in Abdo et
al. (2010), the young pulsars looks to still follow this relation
with a large scattering factors of more than 10 but millisecond
pulsars follow a different relation (see Fig. 6 of Abdo et al.
2010). This correlation was used to estimate some gamma-selected
pulsars (Saz Parkinson et al. 2010). Moreover, the relation of
$L_\gamma\propto \dot E^{1/2}$ may not be intrinsic, for the young
gamma-ray pulsars in Fig. 6 of Abdo et al. (2010), we find a
fitting function of $L_\gamma \propto \dot E^{0.7}$.

Gamma-ray emission efficiency ($\eta=L_\gamma/\dot E$) is an
important parameter in gamma-ray pulsars, which varies for
different populations of pulsars. In this paper we study the
relations of gamma-ray emission efficiency versus some pulsar
parameters: spin period, age, magnetic field at light cylinder,
and three generation order parameters. We will show results of
these relations and good correlations would be pulsar distance
indicators for gamma-selected pulsars.

\section{Gamma-ray emission efficiency versus pulsar parameters}

Gamma-ray emission efficiency is defined as $\eta=L_\gamma/\dot
E$, where the spin-down power $\dot E=4\pi^2 I \dot P P^{-3}$
taking $I=10^{45}$ g cm$^2$, $P$ is the period of pulsar in units
of second. $L_\gamma =4\pi d^2 f_\Omega F_\gamma$, where
$F_\gamma$ is the gamma-ray flux above 100 MeV detected by
Fermi/LAT. The radiation open angle factor $f_\Omega$ is
model-dependent, and may depend on the magnetic inclination and
observer angles, which could be obtained using pulse profile
information (e.g., $f_\Omega\sim 1$ for 8 gamma-ray pulsars
estimated by Watters et al. 2009). For simplicity, we use
$f_\Omega=1$ in this paper similar to Abdo et al. (2010). In the
gamma-ray pulsar catalog given by Abdo et al. (2010), 21
radio-selected young pulsars and 7 gamma-selected pulsars have the
distance measurements or estimation. However, gamma-ray emission
efficiency of some gamma-selected pulsars are higher than 1, the
maximum radiation efficiency in physics, suggesting overestimation
of the distance for some gamma-selected pulsars. Millisecond
pulsars may have different properties from young pulsars, so we do
not consider 8 millisecond pulsars in the catalog here. Finally,
we use these 21 radio-selected pulsars for the analysis in this
section. The efficiency $\eta$ distributes from $0.1\%$ (like Crab
pulsar) to near $100\%$.

We will first show the relations between $\eta$ versus three well
known pulsar parameters: $P$, $\tau$ and $B_{\rm LC}$. Most
importantly, we have introduced the generation order parameters
for pulsars (see details in Wang \& Zhao 2004 and therein
references) which can be used to describe gamma-ray properties in
pulsars. In \S 2.2, the relations between $\eta$ versus three
generation order parameters will be studied.

\subsection{$\eta$ versus $P$, $\tau$ and $B_{\rm LC}$}

In Fig. 1, we plot diagrams of $\eta$ versus $P$, $\tau$ and
$B_{\rm LC}$ for 21 young gamma-ray pulsars, respectively.
$\tau=P/\dot P$ is the pulsar's characteristic age, $B_{\rm
LC}\sim 2.94\times 10^8(\dot P P^{-5})^{1/2}$ is the magnetic
field at the light cylinder ($R_{\rm LC}=cP/2\pi$).

In the diagram of $\eta-P$, the data points of spin period are
scattering, no significant correlation is found. But $\eta$ seems
to have the correlations with the other two pulsar parameters: age
and the magnetic field at the light cylinder. The linear function
is used to fit the correlations (solid lines in Fig. 1):

\beq \log\eta = -(4.73\pm 0.31) + (0.68\pm 0.08)\log \tau \enq
with a standard deviation of $\sigma \sim 1.63$ and the
probability value ($p$-value for t-test) of $1.09\times 10^{-4}$;

\beq \log\eta = (2.23\pm 0.32) - (0.88\pm 0.10)\log B_{\rm LC}
\enq with $\sigma \sim 1.56$, and a $p$-value of $2.90\times
10^{-5}$.

The gamma-ray efficiency generally become higher with older age
and smaller $B_{\rm LC}$. From the evaluation of the standard
deviation values $\sigma$ and $p$-values, the relation of
$\eta-B_{\rm LC}$ is the better one.

\begin{figure*}
\includegraphics[angle=0,width=17cm]{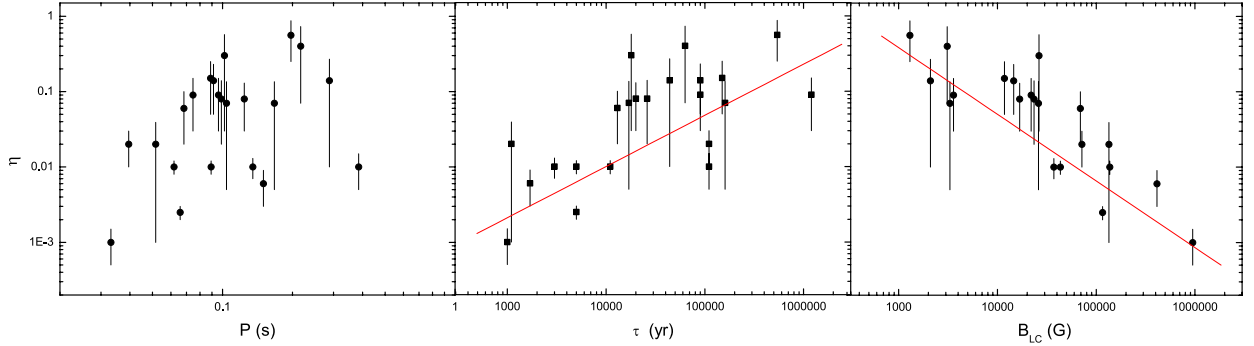}
\caption{Gamma-ray emission efficiency $\eta$ of 21 young
gamma-ray pulsars versus three pulsar parameters: spin period $P$,
age $\tau$ and the magnetic field at light cylinder $B_{\rm LC}$.
$\eta$ shows the correlations with two pulsar parameters $\tau$
and $B_{\rm LC}$, and the solid lines display the best fitting
functions. See the text for details.}
\end{figure*}

\begin{figure*}
\includegraphics[angle=0,width=17cm]{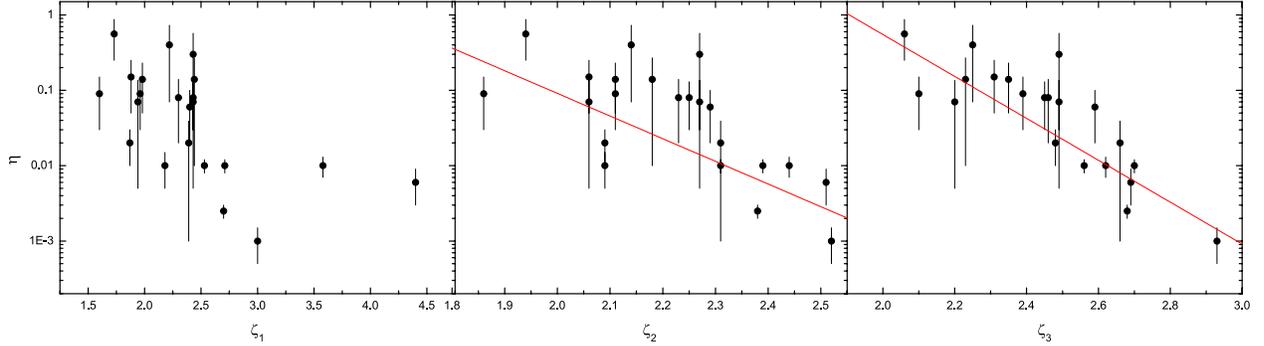}
\caption{Gamma-ray emission efficiency $\eta$ of 21 young pulsars
versus three generation order parameters $\zeta_{1-3}$ in
gamma-ray pulsars. $\eta$ has no significant correlation with
$\zeta_1$ but has the correlation with $\zeta_2$ and $\zeta_3$,
suggesting that the magnetic field dominates the gamma-ray
absorption in cascade processes. The sold lines show the best
fitting function. See the text for details. }
\end{figure*}

\subsection{$\eta$ versus generation order parameters}

The concept of generation is provided to describe pair cascade
processes in gamma-ray pulsars (Zhao et al. 1989, Lu \& Shi 1990).
Based on Ruderman-Sutherland scenario (Ruderman \& Sutherland
1975), passing through the polar gap, $e^+/e^-$ are accelerated to
a high energy with typical Lorentz factor $\gamma_1=6.0\times 10^7
P^{1/14}\dot{P}_{15}^{-1/14}$, where $\dot{P}_{15}$ the derivative
in units of $10^{-15}$s\ s$^{-1}$. These first generation
particles will move along the curved magnetic field lines and emit
high energy curvature radiation (the first generation photons)
with photon energy typically at \beq E_1={3\over 2}{\hbar c\over
R_c}\gamma_1^3\approx 3.2\times
10^{10}P^{-2/7}\dot{P}_{15}^{-3/14}{\rm eV}, \enq where
$R_c\approx 1.8\times 10^7 P^{1/2}$cm is the curvature radius of
field line here. These primary photons could be converted into
secondary $e^+/e^-$ pairs in both open and closed magnetic field
line regions near the neutron star surface due to magnetic pair
creation (Halpern \& Ruderman 1993). And the condition of these
photons to create $e^+/e^-$ pairs is (Sturrock 1971; Ruderman \&
Sutherland 1975)  \beq {E_1\over 2m_e c^2}{B(r_s)\over B_c}\approx
{1\over 15}, \enq where $B(r_s)$ is the local magnetic field at
the position of $r_s$, and $B_c=m_e^2c^3/e\hbar =4.14\times
10^{13}$ G is the critical magnetic field. These $e^+/e^-$ can
emit the second generation photons through synchrotron radiation
with a characteristic energy $E_2$. If $E_2$ is high enough, the
further $e^+/e^-$ pairs (the third generation) can be produced
under the condition similar to Eq. (4), \beq {E_2\over 2m_e
c^2}{B(r_s)\over B_c}\approx {\chi_0\over 15}, \enq where
$\chi_0/15\sim 1/9-1/12$ (Sturrock 1971). Then pair cascade
processes occur.


Concerning this idea, Lu et al. (1994) introduced the generation
order parameter (GOP) to characterize a pulsar. They considered
the conversion of high energy photons into $e^+/e^-$ pairs through
electric fields, and defined the first GOP as \beq
\zeta_1=1+{1-(11/7){\rm log}P+(4/7){\rm log}\dot{P}_{15}\over
3.56-{\rm log}P-{\rm log}\dot{P}_{15}}. \enq

Wei et al. (1997) considered absorption of high energy photons by
the effect of both magnetic and electric fields, define the second
GOP as, \beq \zeta_2=1+{0.8-(2/7){\rm log}P+(2/7){\rm
log}\dot{P}_{15}\over 1.3}. \enq

The concept of generation was initially considered in the scheme
that the $\gamma$-ray photons is absorbed and conversed into
$e^+/e^-$ through only magnetic fields (Zhao et al. 1989), so we
defined the third GOP based on the magnetic field absorption
effects as (Wang \& Zhao 2004) \beq \zeta_3=1+{0.6-(11/14){\rm
log}P+(2/7){\rm log}\dot{P}_{15}\over 1.3}. \enq

GOPs are used to describe cascade processes and characterize the
spectral properties of pulsars. If a pulsar can emit gamma-rays,
the GOPs must be larger than 1 (i.e., the first generation
gamma-ray photons must exist). In addition, the GOPs are proved to
be correlated with the gamma-ray photon index: softer gamma-ray
photons with larger GOPs based on the EGRET pulsar sample (Lu et
al. 1994; Wei et al. 1997). Then according to the definition of
GOPs, the first generation pairs emit high energy gamma-rays
(i.e., $>100$ MeV), with larger GOPs, more first generation pairs
are transferred into next generation pairs with lower energy which
emit more soft gamma-rays and X-rays. So given a total emission
rate, efficiency to GeV gamma-rays ($\eta$) becomes lower with
larger GOPs.

In Fig. 2, we plot the diagrams of $\eta$ versus three GOPs
($\zeta_{1-3}$) respectively. $\eta$ has no correlation with
$\zeta_1$ but has correlation with other two GOPs $\zeta_2$,
$\zeta_3$, implying that magnetic fields dominate the absorption
in pair cascade processes, consistent with our previous results (
Wang \& Zhao 2004). These correlations also suggest that GOPs
($\zeta_2$, $\zeta_3$) can describe gamma-ray properties of
pulsars. In Fig. 2, the solid lines show the best fitting
functions for the relations of $\eta-\zeta_2$ and $\eta-\zeta_3$:

\beq \log\eta = (4.98\pm 0.45) - (3.00\pm 0.21)\zeta_2 \enq with
$\sigma \sim 1.69$ and a $p$-value of $1.94\times 10^{-4}$;

\beq \log\eta = (5.49\pm 0.24) - (2.86\pm 0.11)\zeta_3 \enq with
$\sigma \sim 1.34$ and a $p$-value of $1.01\times 10^{-6}$. The
correlation between $\eta-\zeta_3$ is stronger with the smaller
standard deviation and $p$-value.

\section{Possible distance indicators for gamma-ray selected pulsars}

\begin{table}
\caption{The estimated distances of 25 gamma-selected pulsars.
$d_{1-2}$ denotes the distance range calculated from relations of
$\eta-\zeta_3$ and $\eta-B_{\rm LC}$, respectively. $d_3$ is the
estimated distance from other methods with references provided. }
\begin{center}
\scriptsize
\begin{tabular}{l c c c c c c  l}

\hline \hline Pulsar & $P$ & $\dot P$ & $F_\gamma(>100$ MeV) & $d_1$ & $d_2$& $d_3$ & reference  \\
   &  s    &s s$^{-1}$ & erg cm$^{-2}$ s$^{-1}$ &  kpc  & kpc  & kpc  &   \\
\hline
J0007+7303  &   0.316 & 3.61$\times 10^{-13}$ & 3.82$\times 10^{-10}$  & 0.86$^{+0.30}_{-0.32}$    & 1.18$^{+0.72}_{-0.44}$  & 1.4$\pm 0.3$    & Pineault et al. 1993     \\
J0357+32    &   0.444 & 1.20$\times 10^{-14}$ & 6.38$\times 10^{-11}$  & 0.72$^{+0.25}_{-0.29}$    & 0.73$^{+0.51}_{-0.30}$  &     &     \\
J0633+0632  &   0.297 & 7.95$\times 10^{-14}$ & 8.00$\times 10^{-11}$  & 1.26$^{+0.41}_{-0.48}$    & 1.37$^{+0.76}_{-0.60}$  &     &   \\
J0633+1746  &   0.237 & 1.10$\times 10^{-14}$ & 3.38$\times 10^{-9}$   & 0.19$^{+0.07}_{-0.07}$    & 0.17$^{+0.09}_{-0.06}$  &  0.25$^{+0.12}_{-0.06}$   & Faherty et al. 2007  \\
J1418-5819  &   0.111 & 1.70$\times 10^{-13}$ & 2.35$\times 10^{-10}$  & 1.39$^{+0.58}_{-0.57}$    & 1.86$^{+1.09}_{-0.80}$  &  2-5   & Ng et al. 2005 \\
J1459-60    &   0.103 & 2.55$\times 10^{-14}$ & 1.06$\times 10^{-10}$  & 1.76$^{+0.70}_{-0.67}$    & 1.62$^{+0.97}_{-0.69}$  &     &     \\
J1732-31    &   0.197 & 2.62$\times 10^{-14}$ & 2.42$\times 10^{-10}$  & 0.77$^{+0.41}_{-0.35}$    & 0.86$^{+0.49}_{-0.30}$  &     &     \\
J1741-2054  &   0.414 & 1.69$\times 10^{-14}$ & 1.28$\times 10^{-10}$  & 0.59$^{+0.26}_{-0.25}$    & 0.80$^{+0.48}_{-0.29}$  &  0.38$\pm 0.11$   & Camilo et al. 2009   \\
J1809-2332  &   0.147 & 3.44$\times 10^{-14}$ & 4.13$\times 10^{-10}$  & 0.78$^{+0.31}_{-0.31}$    & 0.81$^{+0.48}_{-0.30}$  &  1.7$\pm 1.0$   &  Oka et al. 1999  \\
J1813-1246  &   0.048 & 1.76$\times 10^{-14}$ & 1.69$\times 10^{-10}$  & 2.18$^{+0.71}_{-0.64}$    & 1.56$^{+1.21}_{-0.68}$  &     &     \\
J1826-1256  &   0.110 & 1.21$\times 10^{-13}$ & 3.34$\times 10^{-10}$  & 1.29$^{+0.56}_{-0.44}$    & 1.39$^{+0.86}_{-0.60}$  &     &     \\
J1836+5925  &   0.173 & 1.49$\times 10^{-15}$ & 5.99$\times 10^{-10}$  & 0.32$^{+0.13}_{-0.14}$    & 0.27$^{+0.15}_{-0.09}$  &  $<0.8$   & Halpern et al. 2007    \\
J1907+0602  &   0.107 & 8.68$\times 10^{-14}$ & 2.75$\times 10^{-10}$  & 1.39$^{+0.46}_{-0.40}$    & 1.42$^{+0.95}_{-0.61}$  &     &     \\
J1958+2846  &   0.290 & 2.10$\times 10^{-13}$ & 8.45$\times 10^{-11}$  & 1.54$^{+0.56}_{-0.51}$    & 1.86$^{+1.01}_{-0.78}$  &     &     \\
J2021+4026  &   0.265 & 5.48$\times 10^{-14}$ & 9.76$\times 10^{-10}$  & 0.38$^{+0.20}_{-0.21}$    & 0.46$^{+0.20}_{-0.18}$  &  $1.5\pm 0.5$  & Landecker et al. 1980    \\
J2032+4127  &   0.143 & 1.98$\times 10^{-14}$ & 1.11$\times 10^{-10}$  & 1.32$^{+0.49}_{-0.52}$    & 1.33$^{+0.71}_{-0.50}$  &  $1.6-3.6$   &  Camilo et al. 2009 \\
J2238+59    &   0.163 & 9.86$\times 10^{-14}$ & 5.44$\times 10^{-11}$  & 2.36$^{+0.75}_{-0.70}$    & 2.64$^{+1.36}_{-0.93}$  &     &     \\
J1023-5746  &   0.111 & 3.84$\times 10^{-13}$ & 2.69$\times 10^{-10}$  & 1.77$^{+0.70}_{-0.55}$    & 2.09$^{+0.95}_{-0.88}$  &     &     \\
J1044-5737  &   0.139 & 5.46$\times 10^{-14}$ & 1.03$\times 10^{-10}$  & 1.72$^{+0.60}_{-0.65}$    & 1.86$^{+0.86}_{-0.72}$  &     &     \\
J1413-6205  &   0.110 & 2.78$\times 10^{-14}$ & 1.29$\times 10^{-10}$  & 2.52$^{+0.89}_{-0.92}$    & 1.56$^{+1.12}_{-0.60}$  &     &     \\
J1429-5911  &   0.116 & 3.05$\times 10^{-14}$ & 9.26$\times 10^{-11}$  & 1.84$^{+0.64}_{-0.69}$    & 1.79$^{+0.97}_{-0.70}$  &     &     \\
J1846+0919  &   0.226 & 9.92$\times 10^{-15}$ & 3.58$\times 10^{-11}$  & 1.52$^{+0.55}_{-0.70}$    & 1.44$^{+0.80}_{-0.51}$  &     &     \\
J1954+2836  &   0.093 & 2.12$\times 10^{-14}$ & 9.75$\times 10^{-11}$  & 1.90$^{+0.67}_{-0.80}$    & 1.72$^{+1.10}_{-0.71}$  &     &     \\
J1957+5033  &   0.375 & 7.08$\times 10^{-15}$ & 2.27$\times 10^{-11}$  & 1.22$^{+0.41}_{-0.45}$    & 1.31$^{+0.65}_{-0.37}$  &     &      \\
J2055+2500  &   0.320 & 4.08$\times 10^{-15}$ & 1.15$\times 10^{-10}$  & 0.56$^{+0.19}_{-0.23}$    & 0.61$^{+0.30}_{-0.19}$  &     &      \\

\hline
\end{tabular}
\end{center}

\end{table}

In \S 2, the relations between $\eta$ and six pulsar parameters:
$P$, $\tau$, $B_{\rm LC}$ and three GOPs $\zeta_1$, $\zeta_2$,
$\zeta_3$ are studied. From the values of standard deviation and
$p$-values after fittings, the correlations of $\eta-\zeta_3$ is
stronger than others, and the correlation of $\eta-B_{\rm LC}$
could also be acceptable. In this paper we do not consider the
physical origin in these correlations. These pulsar parameters can
be estimated by two fundamental measurement parameters $P$ and
$\dot P$ which are relatively easily observed. The gamma-ray
emission efficiency is sensitively dependent on distance
measurement which is very difficult at present, specially nearly
impossible for gamma-selected pulsars. With the strong correlation
of $\eta-\zeta_3$, we have a possible way to estimate a reliable
distance for gamma-ray pulsars with only known $P$, $\dot P$ and
$F_\gamma$.

In the catalog of Abdo et al. (2010), 17 gamma-selected pulsars
are listed and most of them have no any distance information. Saz
Parkinson et al. (2010) claimed detections of 8 new gamma-selected
pulsars in blind frequency searches of Fermi LAT data. In Table 1,
we use the distance indicator obtained by $\eta-\zeta_3$
correlation to estimate the distances of the 25 gamma-selected
pulsars. For comparison, we also give the predicted distance
values calculated using the relation of $\eta-B_{\rm LC}$. From
Table 1, we find the evaluated distances ($d_1,\ d_2$) by
$\eta-\zeta_3$ and $\eta-B_{\rm LC}$ correlations are similar,
suggesting that these two distance indicators can be checked by
each other.

\begin{figure}
\includegraphics[angle=0,width=8cm]{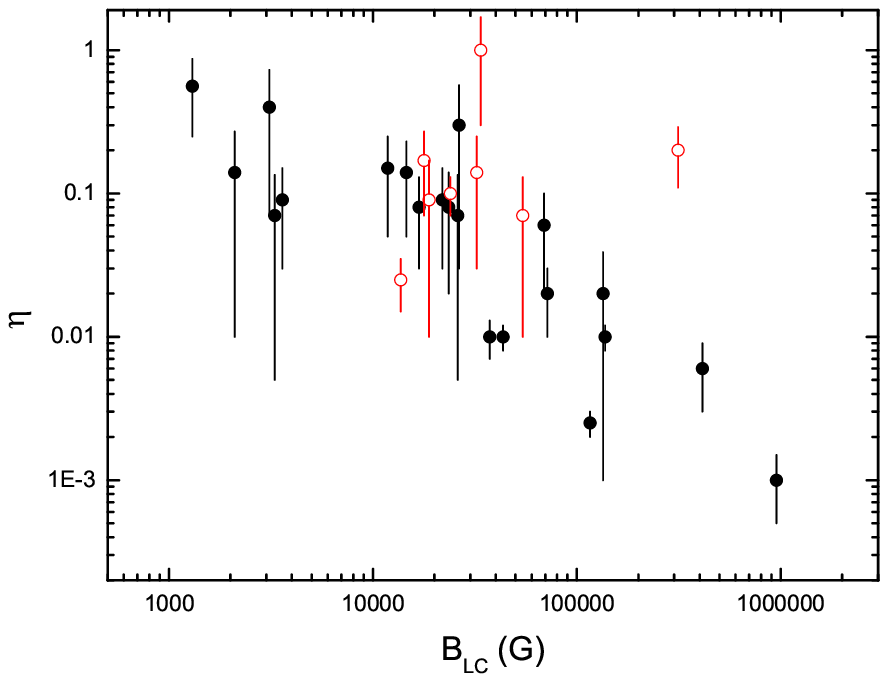}
\includegraphics[angle=0,width=8cm]{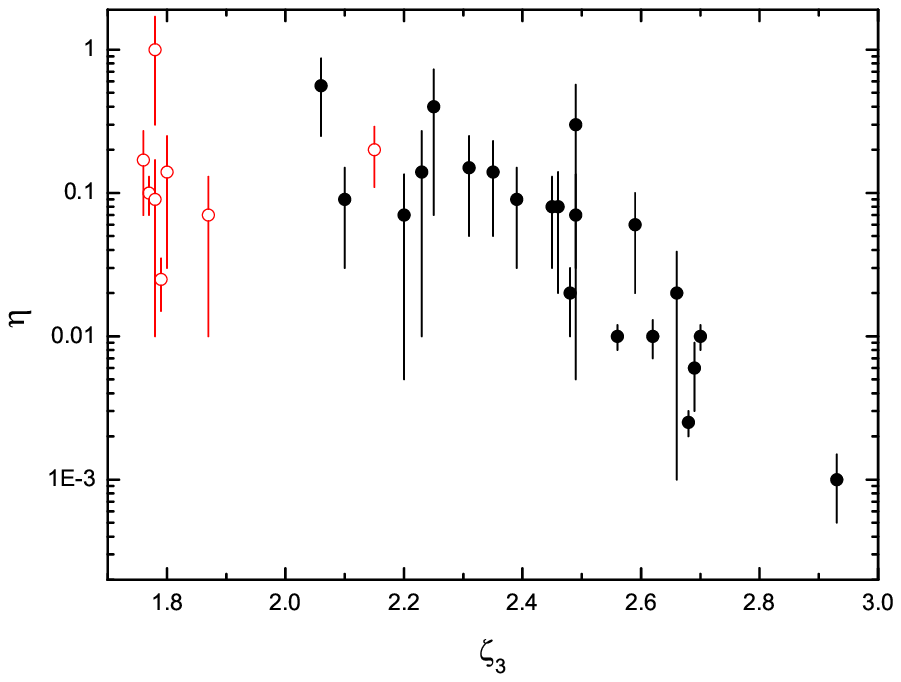}
\caption{Gamma-ray emission efficiency $\eta$ versus $B_{\rm LC}$
and $\zeta_3$ for both 21 young gamma-ray pulsars (solid circles)
and 8 millisecond gamma-ray pulsars (open circles). Millisecond
pulsars still generally follow the relations of $\eta - B_{\rm
LC}$ and $\eta - \zeta_3$ in young pulsars, but they may have a
nearly constant gamma-ray radiation efficiency of $\eta\sim 10\%$.
}
\end{figure}

In Table 1, we also collected the distance information ($d_3$) for
some gamma-selected pulsars from other measurements or
estimations. For the Geminga pulsar, we estimate the distance of
$0.19\pm 0.07$ kpc which is well consistent with the distance
value of $0.25^{+0.12}_{-0.06}$ kpc from the optical trigonometric
parallax measurement (Faherty et al. 2007). For PSR J1836+5925, we
estimate its distance as $\sim 0.3$ kpc (corresponding to an
efficiency of $\sim 55\%$) which is also well below the upper
limits of 0.8 kpc according to its thermal X-ray spectrum (Halpern
et al. 2007). For other gamma-selected pulsars, our estimated
distance values are generally below those from other methods, but
may be more reliable. According to the distance estimated from the
$\eta-\zeta_3$ relation, the gamma-ray efficiency $\eta$ is
general below 1. The estimated efficiency of PSR J2021+4026 is
about $\sim 0.16$ (corresponding to $d\sim 0.4$ kpc) according to
the $\eta-\zeta_3$ relation, compared to $\eta\sim 0.9 - 3.6$
(corresponding to distance of 1 -- 2 kpc) from kinematic model
method on the possible association (Landecker et al. 1980).

\section{Summary and discussion}

\begin{figure*}
\centering
\includegraphics[angle=0,width=9cm]{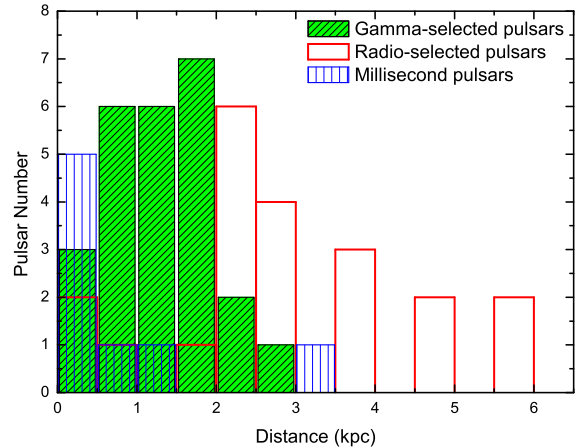}
\caption{The distance distributions of three classes of gamma-ray
pulsars: gamma-selected pulsars, radio-selected pulsars and
millisecond pulsars. The distances of gamma-selected pulsars are
taken from the column $d_1$ of Table 1 according to the distance
indicator of the $\eta-\zeta_3$ correlation. }
\end{figure*}

In this article, we studied the possible correlations between
gamma-ray emission efficiency $\eta$ and 6 pulsar parameters: $P$,
$\tau$, $B_{\rm LC}$ and three generation order parameters
$\zeta_{1-3}$ using 21 young radio-selected gamma-ray pulsars in
Abdo et al. (2010). We find the strong correlation between
$\eta-\zeta_3$. Based on the concept of the GOPs, lager GOPs imply
that more high energy photons are transferred to softer photons
(X-rays). The good correlation of $\eta-\zeta_3$ suggests that
GOPs can describe gamma-ray emission properties of young pulsars,
and the magnetic field absorption effects dominate in pair cascade
processes in pulsar magnetosphere. This intrinsic correlation can
be used to estimate distances for gamma-selected pulsars which
have no any distance information yet. The good correlation of
$\eta-B_{\rm LC}$ is also found, which can be also used as the
other distance indicator in gamma-ray pulsars for a double check.

Millisecond pulsars (MSPs) have not be included in our analysis,
though their distances are generally measured by optical
trigonometric parallax and DM methods. MSPs with much smaller $P$
and $\dot P$ have a much older characteristic age ($\tau\sim 10^9$
yr). The values of $\zeta_1$ and $\zeta_2$ are below 1 or near 1,
making MSPs as non-gamma pulsars if these two GOPS are still
applicable to MSPs. However, in parameter spaces of $B_{\rm LC}$
and $\zeta_3$, MSPs are similar to young pulsars. In Fig. 3, we
plot the diagrams of $\eta-B_{\rm LC}$ and $\eta-\zeta_3$
including both 21 young pulsars and 8 MSPs in the first gamma-ray
pulsar catalog (Abdo et al. 2010). MSPs seems to still follow the
behaviors of young pulsars: higher efficiency with smaller values
of $B_{\rm LC}$ and $\zeta_3$. In the same time, gamma-ray
emission efficiency of MSPs could also be thought to keep constant
$\eta\sim 10\%$ (also see Fig. 6 of Abdo et al. 2010). So MSPs may
have different gamma-ray emission properties from young pulsars,
like multi-pole magnetic field assumption in MSPs (Ruderman 1991;
Zhang \& Cheng 2003), or different emission open angles (taken as
$f_\Omega\sim 0.5$, Fierro et al. 1995). Present discoveries of
MSPs are generally done through radio timing, and the blind search
for MSPs by Fermi/LAT is a very important project in future, but
quite difficult at present specially for MSPs in binaries. Then
the distance indicators of $\eta-B_{\rm LC}$ and $\eta-\zeta_3$
relations could be the secondary way for distance information of
MSPs after trigonometric parallax or DM methods.


The GOP model was originally proposed based on the polar-cap
accelerator scenario. The present Fermi/LAT may support that
gamma-ray emission in pulsars mainly comes from the spatially
extended regions reaching a good fraction of the light-cylinder
radius (e.g., Abdo et al. 2010). The production of the secondary
pairs in polar-cap activity is also different from that in the
scenarios of outer-gap models or slot-gap models (e.g., Cheng, K.
S. et al. 2000; Muslimov \& Harding 2004). Then the new model of
generation order parameters may be developed in the extended
regions from the polar-cap regions to near the light-cylinder
radius. This GOP model would be more complicated but could be
considered in the next work. Anyway, the correlations of
$\eta-\zeta_3$, $\eta-B_{\rm LC}$ for gamma-ray pulsars suggest
that the gamma-ray luminosity may depend on two fundamental pulsar
parameters $P$ and $\dot P$. The function of $P$ and $\dot P$
could well predict gamma-ray emission luminosity, which can be
used to trace the distance of gamma-ray pulsars.

Fig. 4 shows the distance distributions of three classes of
gamma-ray pulsars: gamma-selected pulsars, radio-selected pulsars
and millisecond pulsars. The distances of gamma-selected pulsars
are provided by the distance indicator of the $\eta-\zeta_3$
relation (see Table 1). Gamma-ray loud millisecond pulsars
distribute at a distance peak around 0.3 kpc because MSPs
generally have lower spin-down powers. Gamma-selected young
pulsars distribute at the distance peak of $\sim 1.2$ kpc, while
radio-selected young pulsars distribute at the distance peak of
$\sim 2.5$ kpc. This difference in distance distributions for two
classes of gamma-ray young pulsars may involve further interest.
The nearby unresolved radio-quiet gamma-ray pulsars could
contribute to diffuse gamma-ray background specially for the
high-latitude pulsars located in the Gould Belt (Wang et al.
2005).

Before the Fermi era, only one gamma-selected pulsar Geminga was
known. Now 25 gamma-selected pulsars are discovered, greatly
improving our knowledge of gamma-ray pulsar family. Much more
gamma-selected pulsars would be detected by future deeper sky
surveys of Fermi/LAT. The distance indicators presented in this
paper will give the distance information for gamma-selected
pulsars, which will be helpful for study in gamma-ray emission
properties of this pulsar population. It is still hopefully
expected that more gamma-ray pulsars (young) have trigonometric
parallax measurements or more precise DM model, which can check
the validity of the distance indicators (i.e., $\eta-\zeta_3$,
$\eta-B_{\rm LC}$), and improve the distance indicators in
gamma-ray pulsars.

\begin{acknowledgements}

We are grateful to Han, J.L. and Song, L.M. for the helpful
discussion. This work is supported by the National Natural Science
Foundation of China under grants 10803009, 10833003, 11073030.

\end{acknowledgements}


\begin{thebibliography}{}

\bibitem[]{aa10} Abdo, A. A. et al. 2010, ApJS, 187, 460
\bibitem[]{bw97} Becker, W. \& Tr\"umper, J. 1997, A\&A, 326, 682
\bibitem[]{ca06} Camilo, F. et al. 2006, ApJ, 637, 456
\bibitem[]{ca09} Camilo, F., et al. 2009, ApJ, 705, 1
\bibitem[]{ch00} Cheng, K. S., Ruderman, M. \& Zhang, L. 2000,
ApJ, 537, 964
\bibitem[]{cw04} Cheng, K. S., Taam, R. E., \& Wang, W. 2004, ApJ,
617, 480
\bibitem[]{cl02} Cordes, J.M. \& Lazio, T.J.W. 2002, astro-ph/0207156
\bibitem[]{fa07} Faherty, J., Walter, F. M., \& Anderson, J. 2007, Ap\&SS, 308, 225
\bibitem[]{fe95} Fierro, J. M., et al. 1995, ApJ, 447, 807
\bibitem[]{go03} Gotthelf, E. V. 2003, ApJ, 591, 361
\bibitem[]{hl07} Halpern, J. P., Camilo, F., Gotthelf, E. V. 2007,
ApJ, 668, 1154
\bibitem[]{hr93} Halpern, J. P., \& Ruderman, M.A. 1993, ApJ, 415,
286
\bibitem[]{jo96} Johnston, S. et al. 1996, MNRAS, 279, 1026
\bibitem[]{ke08} Keith, M. J. et al. 2008, MNRAS, 389, 1881
\bibitem[]{la80} Landecker, T. L., Roger, R. S., \& Higgs, L. A. 1980, A\&AS, 39, 133
\bibitem[]{lo06} Lommen, A. N. et al. 2006, ApJ, 642, 1012
\bibitem[]{lu90} Lu, T. \& Shi, T.Y. 1990, A\&A, 231, L7
\bibitem[]{lu94} Lu, T., Wei, D.M., Song, L.M. 1994, A\&A, 290, 815
\bibitem[]{mk04} Muslimov, A. G., Harding, A. K. 2004, ApJ,
606,1043
\bibitem[]{ng05} Ng, C.-Y., Roberts, M. S. E., \& Romani, R. W. 2005, ApJ, 627, 904
\bibitem[]{ok99} Oka, T. et al. 1999, ApJ, 526, 764
\bibitem[]{pi93} Pineault, S. et al. 1993, AJ, 105, 1060
\bibitem[]{po02} Possenti, A. et al. 2002, A\&A, 387, 993
\bibitem[]{ra10} Ray, P. S. \& Saz Parkinson, P. M., 2010,
arXiv:1007.2183
\bibitem[]{ro93} Roberts, D. A. et al. 1993, A\&A, 274, 427
\bibitem[]{ro05} Romani, R. W. et al. 2005, ApJ, 631, 480
\bibitem[]{rs75} Ruderman, M.A. \& Sutherland, P.G. 1975, ApJ, 196, 51
\bibitem[]{ru91} Ruderman, M.A. 1991, ApJ, 366, 261
\bibitem[]{sa98} Saito, Y. 1998, Ph.D.  Thesis, Univ. of Tokyo
\bibitem[]{sp10} Saz Parkinson, P. M. et al. 2010, ApJ, 725, 571
\bibitem[]{st71} Sturrock, P. A. 1971, ApJ, 164, 529
\bibitem[]{th99} Thompson, D. J., et al. 1999, ApJ, 516, 297
\bibitem[]{th01} Thompson, D. J. 2001, in AIP Conf. Series, Vol. 558, High Energy Gamma-Ray
Astronomy: International Symposium, ed. F. A. Aharonian \& H. J.
Volk, 103
\bibitem[]{ww04} Wang, W. \& Zhang, Y. 2004, ApJ, 601, 1038
\bibitem[]{ww05} Wang, W., Jiang, Z. J., Pun, C. S. J., Cheng, K.
S. 2005, MNRAS, 360, 646
\bibitem[]{ww09} Wang, W. 2009, RAA, 9, 1241
\bibitem[]{wa09} Watters, K. P. et al. 2009, ApJ, 695, 1289
\bibitem[]{we97} Wei, D.M., Song, L.M., Lu, T. 1997, A\&A, 323, 98
\bibitem[]{zc03} Zhang, L., \& Cheng, K. S. 2003, A\&A, 398, 639
\bibitem[]{zh89} Zhao, Y. et al. 1989, A\&A, 223, 147

\end{thebibliography}
\end{document}